\documentclass[reprint,superscriptaddress,nofootinbib,amsmath,amssymb,aps,pra]{revtex4-1}

\usepackage{bm}

\newcommand{\Ket}[1]{\left|#1\rangle\right.}

\newcommand{\Integers}{\mathbb{Z}}

\newcommand{\ntot}{n}
\newcommand{\Itot}{I}

\newcommand{\mc}{\mu_c}
\newcommand{\mr}{\mu_r}
\newcommand{\la}{\langle}
\newcommand{\ra}{\rangle}

\newcommand{\cT}{\mathcal{T}}

\newcommand{\da}{\dagger}

\newcommand{\Op}[1]{\hat{#1}}

\newcommand{\oL}{\Op{L}}
\newcommand{\oT}{\Op{T}}
\newcommand{\oP}{\Op{P}}
\newcommand{\oH}{\Op{H}}

\newcommand{\oU}{\Op{U}}

\newcommand{\oK}{\Op{K}}

\newcommand{\diff}{\mathrm{d}}

\DeclareMathOperator{\lcm}{lcm}
\DeclareMathOperator{\tr}{tr}

\usepackage[english]{babel}
\usepackage[utf8x]{inputenc}
\usepackage[T1]{fontenc}

\usepackage{amsmath}
\usepackage{amsfonts}
\usepackage{graphicx}
\usepackage[colorlinks=true, allcolors=blue]{hyperref}
\usepackage{color}
\usepackage{natbib}

\begin{document}

\title{Quantum gears from planar rotors}
\author{Zheng Liu}
\affiliation{Centre for Quantum Technologies, National University of Singapore, 3 Science Drive 2, Singapore 117543, Singapore}
\author{Joshua Leong}
\affiliation{Department of Physics, National University of Singapore, 2 Science Drive 3, Singapore 117542, Singapore}
\author{Stefan Nimmrichter}
\email[Corresponding author: ]{cqtsn@nus.edu.sg}
\affiliation{Centre for Quantum Technologies, National University of Singapore, 3 Science Drive 2, Singapore 117543, Singapore}
\author{Valerio Scarani}
\affiliation{Centre for Quantum Technologies, National University of Singapore, 3 Science Drive 2, Singapore 117543, Singapore}
\affiliation{Department of Physics, National University of Singapore, 2 Science Drive 3, Singapore 117542, Singapore}

\date{\today} 

\begin{abstract}
We investigate the dynamics of interacting quantum planar rotors as the building blocks of gear trains and nano-machinery operating in the quantum regime. Contrary to a classical hard-gear scenario of rigidly interlocked teeth, we consider the coherent contact-less coupling through a finite interlocking potential and study the transmission of motion from one externally driven gear to the next as a function of the coupling parameters and gear profile. 
The transmission is assessed in terms of transferred angular momentum and transferred mechanical work.
We highlight the quantum features of the model such as quantum state revivals in the interlocked rotation and interference-enhanced transmission, which could be observed in prospective rotational optomechanics experiments.
\end{abstract}

\maketitle

\section{Introduction}\label{sec:introduction}

The quantum and classical dynamics of microscopic rotors and orientational degrees of freedom has intrigued both chemists and physicists for a long time. Apart from the historic relevance of rotations in the description of atomic and molecular spectra, coupled rotors have been employed as model systems for synchronization \cite{Jain1984}, and periodically kicked rotors played an important part in the understanding of classical chaos and its quantum signatures such as dynamical localization \cite{Casati1979,Fishman1982,Haake2001,Wimberger2014}.

The dynamics and coupling of molecular rotors is widely studied in physical chemistry, e.g.~as components of molecular machines \cite{Schliwa2003,Kottas2005,Kay2007,Michl2009}, albeit in the classical regime. However, with the rapid experimental progress in the recently established field of rotational optomechanics \cite{Shi2013,Yin2013,Muller2015,Kuhn2015,Shi2016,Hoang2016,Xu2017,Kuhn2017,Kuhn2017a,Ahn2018}, nanomechanical realizations of quantum rotors may soon be available. They could act as the thermally driven flywheels in autonomous quantum models for rotor heat engines \cite{Levy2016,Roulet2017,Seah2018,Roulet2018,Seah2018a,Fogedby2018}.

Here we study mutually coupled quantum planar rotors, e.g.~nanoparticles rotating in a fixed plane or particles on a ring, as a direct realization of quantum gears \cite{MacKinnon2002}. In particular, we assess their coherent quantum dynamics and the transmission of motion from one gear to the next, as well as its dependence on the gear profile  and on the interaction potential that determines the interlocking between the gears' ``teeth''. 

To highlight the nonclassical features in our model, we consider the deep quantum limit of two gears prepared in their interlocked ground state and the controlled excitation of motion by a short sequence of angular momentum kicks applied to one of them. Our theoretical model for the gears will be introduced in Sec.~\ref{sec:preliminary}, before we apply it to analyze the transmission of angular momentum and useful mechanical energy from the kicked gear to the other in Sec.~\ref{sec_QM2}. The useful part of the energy is what can be extracted from the second gear (or further transmitted) as mechanical work; it is measured in terms of the so-called ergotropy \cite{Allahverdyan2004,Goold2016}.
We will discuss the influence of angular momentum quantization and quantum tunneling on the dynamics, and the resulting effects of quantum state revivals in the synchronous gear motion and interference-enhanced transmission of angular momentum for certain kick strengths. 

\section{Theory of two coupled gears}\label{sec:preliminary}

\begin{figure}
\centerline{\includegraphics[width=6cm]{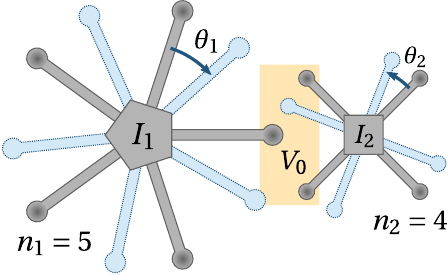}}
\caption{\label{fig:sketch}Exemplary sketch of two coupled gears, modeled as planar rotors with angle coordinates $\theta_{1,2}$, moments of inertia $I_{1,2}$ and teeth numbers $n_{1,2}$. The teeth interact through a periodic potential of strength $V_0$ that is invariant under combined rotations for which $n_1\theta_1 - n_2\theta_2=$const (blue shades). }
\end{figure}

We consider two coupled gears represented by planar quantum rotors with moments of inertia $I_1,I_2$ and integer numbers of teeth $n_1 , n_2$, as sketched in Fig.~\ref{fig:sketch}. Given the angle coordinates $\theta_1, \theta_2$ and the corresponding angular momenta $L_1, L_2$, the coupled two-gear system will be described by the Hamiltonian \cite{MacKinnon2002}
\begin{equation}
H=\frac{L_1^2}{2I_1}+\frac{L_2^2}{2I_2}-V_0\,u(n_1\theta_1-n_2\theta_2). \label{eqn:H}
\end{equation}
Here we model the coupling in terms of an attractive interlocking potential of depth $V_0 > 0$ and a generic $2\pi$-periodic function $0\leq u(x)\leq 1$. Evaluated at $x=n_1\theta_1 - n_2 \theta_2$, the latter sets the angular dependence of the interlocking, which does not distinguish between the individual teeth of each gear. 

Besides these requirements, the potential may often also have a definite parity, either even $u(-x)=u(x)$ or odd $u(-x)=-u(x)$. In our numerical case studies, we shall use the even function 
\begin{equation}
u(x)=\frac{1 + \cos x}{2}. \label{eq:u_pendulum}
\end{equation} 
Experimental realizations of such gear couplings can be envisaged on the level of rotating nanoparticles. For example, the case $n_1=n_2=1$ can be implemented using altitudinal molecular dipole chains described in \cite{Jonge2004}. Induced dipole-dipole coupling between two anisotropic dielectric objects rotating around the same axis yields a potential energy proportional to the polarization squared, $\propto \cos^2(\theta_1-\theta_2)$, which amounts to $n_1=n_2=2$. Experimental candidate systems for these objects could be optically levitated nanorods currently under experimental investigation \cite{Kuhn2015,Hoang2016,Xu2017,Kuhn2017,Kuhn2017a,Ahn2018}. Higher teeth numbers arise, for instance, by studying the motion of charged particles on coplanar rings \cite{Averin2018}.

\subsection{Classical gear motion}
\label{sec:classical}
A good intuition on the expected average gear dynamics can be gained by looking at the classical problem first. Here a coordinate transformation reveals that the coupled gear motion is comprised of a free ``center-of-mass'' rotation of the interlocked gears and a ``relative'' pendulum motion against the interlocking potential.

Explicitly, given that the potential depends only on the linear combination $n_1\theta_1-n_2\theta_2$ of the gear angles, we can define new variables for the center-of-mass and the relative rotations, 
\begin{equation}
\theta_c = \frac{n_2I_1}{\ntot \Itot} \theta_1 + \frac{n_1I_2}{\ntot \Itot} \theta_2 , \quad \theta_r = \frac{n_1}{\ntot} \theta_1 - \frac{n_2}{\ntot} \theta_2,
\label{eqn:change_of_variable_theta}
\end{equation}
where $\Itot = (I_1 + I_2)/2$ and $\ntot = n_1+n_2$. The conjugate angular momenta and the corresponding moments of inertia are
\begin{eqnarray}
L_c &=& \frac{n_2 I_c}{\ntot \Itot} L_1 + \frac{n_1 I_c}{\ntot \Itot} L_2 , \quad I_c = \frac{\ntot^2 \Itot^2}{n_1^2 I_2 + n_2^2 I_1}, \nonumber \\
L_r &=& \frac{n_1 I_r}{\ntot I_1}L_1 - \frac{n_2 I_r}{\ntot I_2}L_2, \quad I_r = \frac{\ntot^2 I_1 I_2}{n_1^2 I_2 + n_2^2 I_1}.
\label{eqn:change_of_variable_Lcr}
\end{eqnarray}
With the above change of variables, the Hamiltonian splits into two uncoupled parts for the free center-of-mass rotation and the relative motion,
\begin{equation}
H=H_c+H_r=\frac{L_c^2}{2I_c}+\frac{L_r^2}{2I_r}-V_0\,u(\ntot \theta_r). \label{eqn:H_cr}
\end{equation}
The momentum $L_c$ is conserved, and the classical equations of motion follow as $\theta_c(t) = \theta_c(0) + L_c t/I_c$ and
\begin{equation}
\dot{\theta}_r(t) = \frac{L_r (t)}{I_r}, \quad \dot{L}_r(t) = \ntot V_0 u'[\ntot \theta_r (t)].
\label{eqn:L_r_eom}
\end{equation}
For the cosine potential \eqref{eq:u_pendulum}, the relative coordinates describe the motion of a pendulum. 

The interlocking between the classical gears and the transmission of motion from one gear to the other can be understood as a simple threshold behavior. The two gears are in a perfectly interlocked state, and moving synchronously like ideal hard gears, as long as $\theta_r$ is at the minimum of the potential and $L_r=0$. Now suppose that Gear 1 suddenly receives an impulse that imparts the momentum $\Delta L_1$. According to \eqref{eqn:change_of_variable_Lcr}, a part of it will simply raise the velocity of the interlocked rotation, while the remainder induces relative oscillations in the interlocking potential. However, once the relative momentum exceeds the interlocking threshold 
\begin{equation}
L_r = \sqrt{2 I_r V_0}, \label{eq:interlockThresh}
\end{equation}
the relative motion is no longer bounded by turning points of zero kinetic energy; the angle $\theta_r (t)$ will increase (decrease) monotonically and Gear 2 will fall behind Gear 1.

\subsection{Periodic boundary conditions}
\label{sec:QM1}

For quantum gears, the joint wave function must be strictly $2\pi$-periodic in both $\theta_1$ and $\theta_2$, i.e.~invariant under the group of symmetry transformations
\begin{equation}
\cT_{k_1, k_2}: \, (\theta_1,\theta_2) \mapsto (\theta_1 + 2\pi k_1,\theta_2 + 2\pi k_2), \quad k_1,k_2 \in\Integers. \label{eq:SymTrafo}
\end{equation}
This implies that the eigenvalues of the angular momentum operators $\oL_1$ and $\oL_2$ are integer multiples of $\hbar$, denoted by the quantum numbers $m_1,m_2\in\Integers$. The corresponding eigenstates, $\oL_{1,2} \Ket{m_1,m_2}=m_{1,2}\hbar\Ket{m_1,m_2}$, satisfy
\begin{equation}
\la \theta_1,\theta_2|m_1,m_2 \ra = \frac{1}{2\pi}e^{im_1 \theta_1 + im_2\theta_2} .
\end{equation}
Note that the angle operators $\hat{\theta}_j$ conjugate to $\oL_j$ and the canonical commutation relations cannot be defined in a straightforward manner \cite{Carruthers1968}. In fact, we should only speak of functions such as $u(n_1 \hat{\theta}_1 - n_2 \hat{\theta}_2)$, which are $2\pi$-periodic in both rotor angles and thus well defined in the operator sense. They can be Fourier-expanded in terms of the unitary kick operators
\begin{equation}
\oK_{\ell_1,\ell_2} = \exp \left[ i \ell_1 \hat{\theta}_1 + i \ell_2 \hat{\theta}_2 \right], \quad \ell_1,\ell_2 \in\Integers, \label{eq:kickOperator}
\end{equation}
which describe a displacement of $\oL_1$ and $\oL_2$ by $\ell_1\hbar$ and $\ell_2 \hbar$, respectively.

We can now transform to relative and center-of-mass coordinates according to \eqref{eqn:change_of_variable_theta} and \eqref{eqn:change_of_variable_Lcr}, but the discrete  spectra of $\oL_{1,2}$ and the strict periodicity in the angle representation will generally imply nontrivial spectra and combined boundary conditions for the new coordinates. 
In momentum representation, the operators $\oL_1,\oL_2,\oL_c,\oL_r$ are mutually commuting and so the quantum numbers $(m_1,m_2)$ determine the spectrum of $(\oL_c, \oL_r)$. In general, their eigenvalues $(\mc , \mr )$ in units of $\hbar$ will not be integers, but take discrete values 
\begin{equation}
\mc = \frac{I_c}{\ntot \Itot} (n_2 m_1 + n_1 m_2), \,\,\, \mr = \frac{I_c}{\ntot \Itot^2} (n_1I_2m_1 - n_2I_1m_2). \label{eq:mc_mr}
\end{equation}
The corresponding product states $\Ket{m_1,m_2}$ and $\Ket{\mc ,\mr }$ denote the same simultaneous eigenstate. 

In the general case where the inertia ratio $I_1/I_2$ is not a simple rational number, \eqref{eq:mc_mr} reveals an important distinction between the center-of-mass and the relative degree of freedom. While the latter may have an incommensurate spectrum of momentum eigenvalues and thus no periodicity in the coordinate $\theta_r$, the $\mc$-values are all integer multiples of the same real number $I_c/\ntot \Itot$. The resulting periodicity in the center-of-mass coordinate follows by writing $\mc$ in a fully reduced form, 
\begin{equation}
\mc = \frac{M_1 m_1 + M_2 m_2}{\nu}, \quad \nu = \frac{M_1^2 I_1 + M_2^2 I_2}{(M_1+M_2) \Itot}. \label{eq:mc_reduced}
\end{equation}
Here, the reduced integer factors $M_{1,2}$ are given in terms of the least common multiple or greatest common divisor of the teeth numbers,
\begin{equation}
M_{1,2} = \frac{\lcm (n_1,n_2)}{n_{1,2}} = \frac{n_{2,1}}{\gcd(n_1,n_2)}. \label{eq:M12}
\end{equation}
The term $\nu$ denotes the period of the center-of-mass coordinate $\theta_c$ in units of $2\pi$. Indeed, one of the symmetry transformations \eqref{eq:SymTrafo} yields
\begin{equation}
\cT_{M_1,M_2}: \, (\theta_c,\theta_r) \mapsto (\theta_c+2\pi\nu,\theta_r). \label{eq:com_periodicity}
\end{equation}
However, this separate center-of-mass periodicity is a derived constraint, since it follows from multiple applications of the basic symmetries $\cT_{1,0}$ and $\cT_{1,0}$. These dictate that the two-gear state be invariant under
\begin{eqnarray}
\cT_{1,0}: \, (\theta_c,\theta_r) &\mapsto& \left( \theta_c + \frac{ 2\pi n_2 I_1}{\ntot \Itot}, \theta_r + \frac{ 2\pi n_1}{\ntot} \right), \nonumber \\
\cT_{0,1}: \, (\theta_c,\theta_r) &\mapsto& \left( \theta_c + \frac{ 2\pi n_1 I_2}{\ntot \Itot}, \theta_r - \frac{ 2\pi n_2}{\ntot} \right). \label{eq:basicSymTrafos}
\end{eqnarray}
This joint boundary condition in the $(\theta_c,\theta_r)$-representation implies that the separation ansatz for the two-gear state, $\psi (\theta_c,\theta_r) = \psi_c(\theta_c) \psi_r(\theta_r)$, will not always be appropriate and linear combinations of product states must be considered.

At the same time, the general center-of-mass periodicity \eqref{eq:com_periodicity} implies that there are \emph{quantum revivals} in the coupled gear motion. Since the interlocked gear rotation is decoupled from the relative motion and free, with its energy eigenvalues given by $\hbar^2 \mc^2/2I_c$, the center-of-mass state will rephase at multiples of the revival time
\begin{equation}
\tau_c = \frac{4\pi I_c}{\hbar} \nu^2 = \frac{4\pi (M_1^2 I_1 + M_2^2 I_2)}{\hbar}.
\end{equation}
This constitutes an observable quantum feature \cite{Seideman1999,Poulsen2004,Seideman2005,Stickler2018a}.

A separate periodic boundary condition, and regular spectrum of momentum quanta, in the relative coordinate $\theta_r$ exists only for commensurate gears. Specifically, in the case of equal moments of inertia, $I_1,I_2=I$, both the center-of-mass and the relative momentum values will be quantized in units of 
\begin{equation}
\frac{1}{\nu} = \frac{M_1+M_2}{M_1^2+M_2^2} = \frac{\ntot \gcd(n_1,n_2)}{n_1^2 + n_2^2}, \label{eq:nu_inv}
\end{equation}
as follows from \eqref{eq:mc_mr}. Hence the two-gear state is $2\pi\nu$-periodic not only in  $\theta_c$, but also in $\theta_r$, as can be seen by applying $\cT_{M_2,-M_1}$.

For the case studies below, we will consider identical gears with $I_1,I_2=I$ and $n_1=n_2$. In this case, we have $\nu=1$, integer values of $\mc$ and $\mr$, and $2\pi$-periodicity in $\theta_c$ and $\theta_r$. However, these constraints are not sufficient. Invariance under the elementary symmetry transformations \eqref{eq:basicSymTrafos} yields the necessary and sufficient joint boundary condition
\begin{equation}
\psi(\theta_c+\pi,\theta_r\pm\pi)=\psi(\theta_c,\theta_r).\label{eqn:cbc_allequal}
\end{equation}
For a valid product wave function $\psi_c (\theta_c) \psi_r (\theta_r)$, both factors must have the same even or odd symmetry under a $\pi$-rotation. 
Another way of putting it is that a valid two-gear state may contain only those basis vectors $|\mu_c,\mu_r\ra$ where the two quantum numbers are either both even or both odd; this ensures that $m_1$ and $m_2$ are integers.

\subsection{Symmetries in the relative coordinate}\label{sec:sub_symmetry}

We have seen that the quantum dynamics of two coupled gears separates into a free interlocked rotation and the relative motion against the interlocking potential. The former is described by a periodic center-of-mass coordinate and it exhibits quantum state revivals. We now discuss the properties of the relative degree of freedom.

The relative Hamiltonian $\oH_r = \oL_r^2/2I_r - V_0 u(\ntot \hat{\theta}_r)$ describes the motion in a periodic potential. The associated symmetry transformations are translations of $\theta_r$ by multiples of $2\pi/n$, represented by the unitary operator $\oT_r = \exp[2\pi i \oL_r /\ntot]$ that commutes with $\oH_r$.

According to Bloch's theorem, the simultaneous eigenfunctions of $\oH_r$ and $\oT_r$, i.e.~the stationary states of relative motion, can be written as
\begin{equation}
\psi_{j,k}(\theta_r) = e^{i k \theta_r}u_{j,k}(\theta_r), \quad u_{j,k} \left(\theta_r + \frac{2\pi}{\ntot} \right) = u_{j,k} (\theta_r). \label{eqn:bloch_eigenfunctions}
\end{equation}
By restricting the Bloch wave number to the first Brillouin zone, $k\in\left(-\frac{\ntot }{2},\frac{\ntot }{2}\right]$, we group the associated energies into separate bands labeled by an additional band index $j \in \mathbb{N}$. The periodicity of the Bloch function $u_k$ allows us to Fourier-expand,
\begin{equation}
\psi_{j,k}(\theta_r)= \sum_{m_r\in\Integers} C_{j,m_r}e^{i(k+m_r \ntot)\theta_r}. \label{eqn:bloch_eigenfunctions_expansion}
\end{equation}
However, not all values of $k$ and $m_r$ will result in a valid relative state. Recall that the two-gear wave function must satisfy periodic boundary conditions, which results in joint constraints for the relative and center-of-mass coordinate. In particular, only those relative momenta $\mu_r = k + m_r \ntot$ are allowed that coincide with the discrete spectrum in \eqref{eq:mc_mr}. For equal moments of inertia, this implies that $\mu_r$ is a multiple of \eqref{eq:nu_inv}, which leaves us with only a few discrete $k$-values per band. For identical gears, we get bands of $\ntot$ energy levels with integer $k=-\ntot/2+1,\ldots, \ntot/2$.

In the case studies below, we employ an interlocking potential of even parity, which increases the symmetry of the problem. Namely, the parity operator $\oP_r = \sum_{\mu_r} |-\mu_r \ra \la \mu_r| $ commutes with $\oH_r$, which means that there exists a simultaneous eigenbasis. 
Since $\oP_r$ flips the sign of the Bloch wave number $k$, this implies that the energy bands are degenerate and the states $\psi_{j,k}$ and $\psi_{j,-k}$ have the same energy. The only non-degenerate states are the ones associated to $k=0$ and $\ntot/2$ in the center and at the boundary of the first Brillouin zone; they have definite parity and zero average momentum, $\la \oL_r \ra = 0$.

\begin{figure}
\centering
\includegraphics[width=\columnwidth]{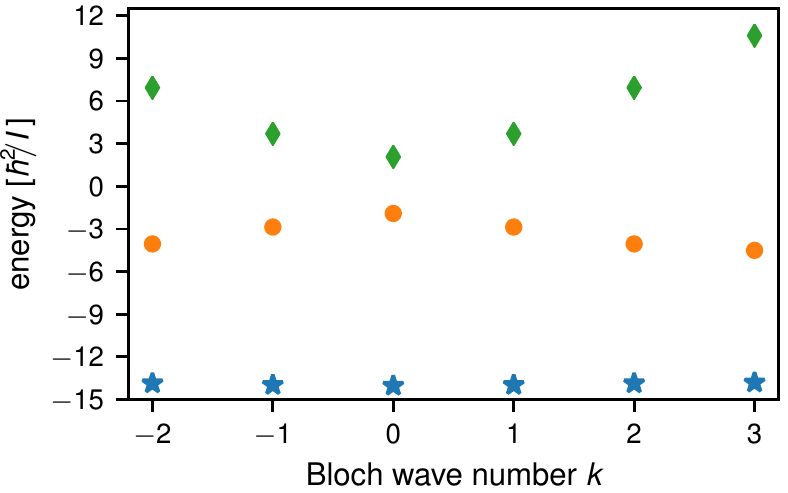}
\caption{The first three energy bands for the relative motion of two identical gears with $n_{1,2}=3$ and $V_0=20.0\hbar^2/I$, plotted against the associated Bloch wave numbers $k=-2..3$ in the first Brillouin zone. The bands are symmetric with degenerate pairs at $k=\pm 1$ and $\pm 2$ and the two non-degenerate states in the center and at the edge of the zone, $k=0$ and $3$. Interlocked gears are only described by the first two bands, the third comprises only unbounded states of positive energy.}
\label{fig:brilluoin_zone_3_3_20}
\end{figure}

As an illustration, we plot in Fig.~\ref{fig:brilluoin_zone_3_3_20} the first three bands in the case of identical gears with $n_{1,2}=3$ teeth and an interlocking depth of $V_0=20.0\ \hbar^2/I$. The first two bands correspond to interlocked gears with negative energies in the relative motion, whereas the third band is beyond the interlocking threshold. 
In the following, we will see that an effective transmission of angular momentum between quantum gears can be achieved beyond interlocking.

\section{Transmission of motion between quantum gears}\label{sec_QM2}

We will now illustrate the quantum behavior of two coupled gears with a case study on the transmission of motion between them. To this end, we consider the ideal quantum scenario where the gears are initially prepared in their interlocked ground state and motion is initiated in a controlled manner by imparting quanta of angular momentum to Gear 1. Our main focus here is to study how much of that externally supplied momentum will be transmitted to Gear 2 over time or on average. We can measure this transmission ratio directly by comparing the angular momentum  Gear 2 acquires over time (or on average) with the total momentum input received by Gear 1. Alternatively, we can look at the ergotropy \cite{Allahverdyan2004,Goold2016}, i.e.~the amount of useful energy that ends up in the reduced state of Gear 2. 

We will first consider the case where the momentum is supplied in a single kick event in Sec.~\ref{sec:sub_single_kick}, which results in a quantum-enhanced transmission for certain kick strengths and in a sequence of smaller kicks. Then we will assess the transmission of ergotropy in Sec.~\ref{sec:sub_ergotropy}, and we will discuss the transmission in a sequence of kick events in Sec.~\ref{sec:sub_multiple_kicks}.

In all instances, we assume equal moments of inertia, and our starting point is the interlocked quantum ground state 
\begin{equation}
\psi (\theta_c, \theta_r, t=0) = \frac{1}{\sqrt{2\pi \nu }} \psi_{1,0} (\theta_r)=\sum_{m_r} c_{m_r}e^{i m_r \ntot \theta_r}, \label{eq:2gearGroundState}
\end{equation}
It is a product of the constant center-of-mass eigenstate ($\mu_c=0$) and the Bloch wave function \eqref{eqn:bloch_eigenfunctions_expansion} associated to the minimum binding energy in the interlocking potential, where the coefficients $c_{m_r}$ are nonzero only for $m_r n \nu \in\Integers$. 

An alternative starting point that rather mimics the behavior of classical gears would be a two-gear state localized in a particular interlocked angle configuration. Unlike the ground state we consider here, such a configuration would not be described by a single stationary product state, but by a linear combination of center-of-mass wave packets combined with Bloch wave functions in the relative coordinate.
The time evolution of the average angular momenta after a kick would then resemble classical gear motion (averaged over a distribution of initial conditions). The remaining quantum features, dispersion and quantum state revivals in the center-of-mass coordinate, would disappear in a scenario of continuous external drive, where a time-dependent potential keeps accelerating Gear 1.

\subsection{Transmission of a single kick}
\label{sec:sub_single_kick}

We consider a single kick event at $t=0$ where a total $\ell \hbar$ of angular momentum is imparted on Gear 1. It can be described by applying a unitary kick operator \eqref{eq:kickOperator} onto the initial state \eqref{eq:2gearGroundState},
\begin{equation}
\oK_{\ell,0} = \exp [i\ell\hat{\theta}_1] = \exp \! \left[\frac{iM_1 \ell \hat{\theta}_c}{\nu} \right] \otimes \exp \!\left[\frac{iM_2 \ell \hat{\theta}_r}{\nu} \right]. \label{eq:kick1}
\end{equation}
This imparts $M_1\ell$ and $M_2 \ell$ quanta to the center-of-mass and the relative momentum, both quantized in units of $\hbar/\nu$. The center-of-mass state remains stationary, but the relative state doesn't, which will lead to time-dependent, typically oscillating, average angular momenta $\la \oL_{1,2} (t) \ra$. To assess the net overall transfer of directed motion to Gear 2, we thus introduce the time-averaged transmission ratio
\begin{equation}
r=\frac{\overline{L_2} }{\ell \hbar}, \quad \overline{L_2} = \lim_{t\to\infty} \frac{1}{t} \int_0^t \diff t' \, \la \oL_2 (t')\ra . \label{eq:transmRatio}
\end{equation}
For a classical benchmark value, suppose the gears are initially at rest in a minimum of the interlocking potential. The kick will leave the gears interlocked as long as the increase in relative momentum does not reach the threshold \eqref{eq:interlockThresh}, i.e.~$\ell \leq \sqrt{2\Itot V_0 (n_1^2+n_2^2)}/n_1 \hbar$. The relative motion then stays bounded and its angular momentum $L_r (t)$ will average out over time, $\overline{L_r} = 0$. What remains is the conserved center-of-mass momentum, which divides into $\overline{L_{1,2}} = n_{2,1} L_c/\ntot$, so that
\begin{equation}
r_{\rm cl} = \frac{n_1 L_c }{\ntot \ell \hbar} = \frac{n_1 n_2}{n_1^2 + n_2^2}. \label{eq:transmRatioBench}
\end{equation}
Beyond threshold, we have $\overline{L_r}>0$ and the transmission ratio of classical gears will eventually vanish as $\ell \to \infty$.

\begin{figure}
\centering
\includegraphics[width=\columnwidth]{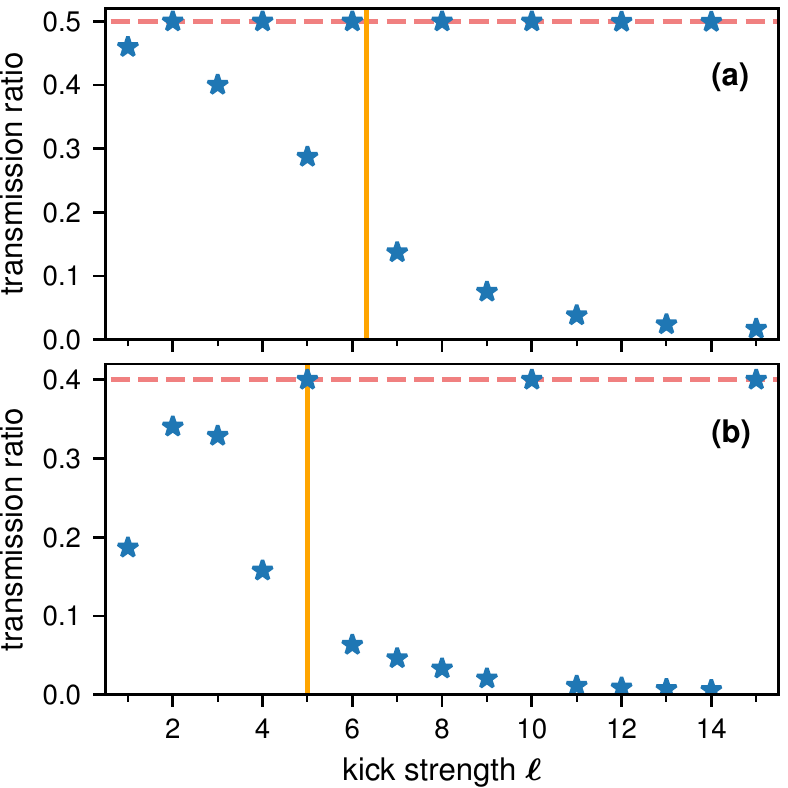}
\caption{Time-averaged transmission ratio of angular momentum between the gears after Gear 1 receives $\ell$ quanta in a single kick. We assume equal moments of inertia $I_{1,2}=I$ for gears with (a) $n_{1,2}=2$ teeth and (b) $n_1=4$ and $n_2=2$. The ideal transmission ratio for perfectly interlocked classical gears is $0.5$ and $0.4$, respectively (dashed lines). Here the gears are initially in the ground state at an interlocking depth $V_0=10.0 \hbar^2/I$. The vertical line marks where the kick supplies enough energy to break interlocking.}
\label{fig:r_single_kick}
\end{figure}

The results for the kicked quantum ground state are different. We find that the transmission ratio can be enhanced to $r_{\rm cl}$ even beyond the classical threshold. Figure~\ref{fig:r_single_kick} shows the ratio as a function of the kick strength $\ell$ for (a) two identical gears with two teeth each, and (b) $n_1=4$ and $n_2=2$. The interlocked benchmark value (red dashed line) is reached at multiples of $2$ or $5$ even when $\ell$ exceeds the classical threshold (vertical line). On the other hand, we also find a reduced transmission below the interlocking threshold for other kick strengths, which can be attributed to quantum tunneling between neighboring minima of the interlocking potential. The transmission ratio can be brought closer to the classical benchmark by increasing the interlocking depth $V_0$.

This quantum enhancement can be explained by the constructive interference of Bloch waves in the relative coordinate. 
There, the kick operator \eqref{eq:kick1} causes a momentum displacement by  
\begin{equation}
\Delta \mu_r = \frac{M_2 \ell}{\nu} = \frac{\ntot n_1 \ell}{n_1^2+n_2^2} \equiv \Delta m_r \ntot + \Delta k, \,\,\, \Delta m_r \in \Integers,
\end{equation}
which may induce transitions from one energy band into several others, but always at a fixed shift $\Delta k$ of the Bloch wave number. 
The enhancement occurs whenever $\Delta \mu_r$ is a multiple of $\ntot/2$, i.e.~for kick strengths $\ell \in \Integers$ that are multiples of $(n_1^2 + n_2^2)/2 n_1$.
The initial ground state ($k=0$) then excites into a (non-stationary) linear combination of Bloch states with a fixed Bloch wave number of either $k=0$ or $k=\ntot/2$ (the latter of which only exists for even $\ntot$).
As each such Bloch state satisfies $\la \oL_r \ra = 0$, the linear combination yields an relative momentum that oscillates around zero and thus averages out over time, $\overline{L_r} = 0$. Hence the interlocked benchmark transmission \eqref{eq:transmRatioBench} is achieved in the long-time average, regardless of whether the occupied energies are negative or positive. 
For identical gears, the enhancement takes place whenever the kick strength $\ell$ is a multiple of $\ntot/2$.

\begin{figure}
\centering
	\includegraphics[width=\columnwidth]{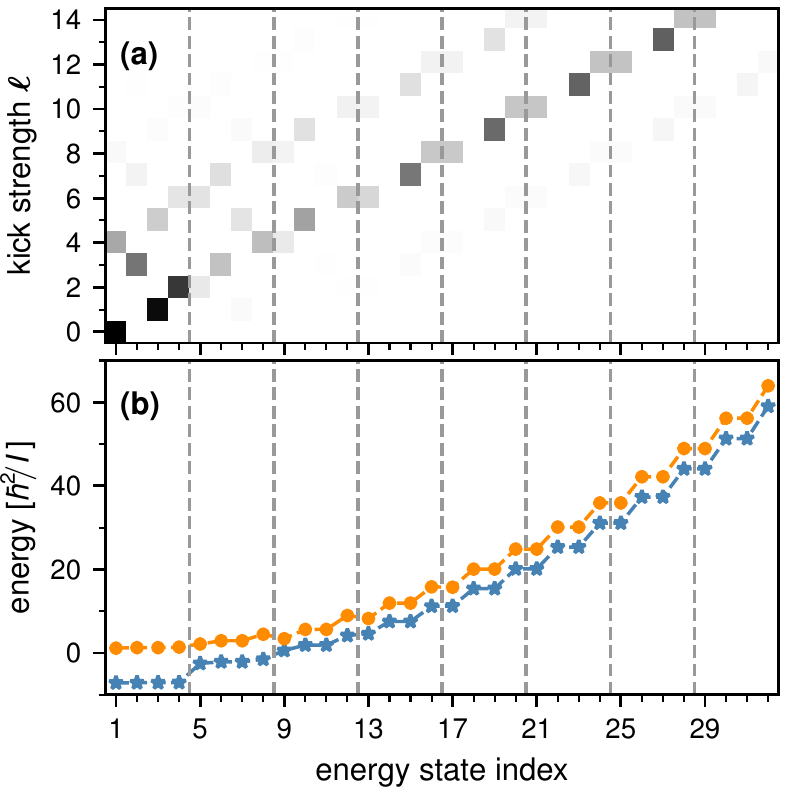}
	\caption{(a) Bloch state occupation probabilities after a single kick on Gear 1 of strength $\ell$. Values from 0 and 1 are mapped to a grey tone from white to black. We use identical gears prepared in the ground state, with $n_{1,2}=2$ and $V_0 = 10 \hbar^2/I$. The Bloch states are labeled by a single index in ascending order of energy, where each block of 4 states forms an energy band. (b) The corresponding spectrum of energy eigenvalues (blue) and average kinetic energies (orange). In high bands beyond interlocking, the relative gear motion is essentially free and most of the energy is kinetic energy.}
\label{fig:kick_energy_spec}
\end{figure}

We illustrate the Bloch state contributions after a single kick event for the simplest case of identical two-teeth gears in Fig.~\ref{fig:kick_energy_spec}. Panel (a) shows the probability to occupy each energy eigenstate of the relative coordinate as a function of the kick strength $\ell$ applied to Gear 1. We assume the gears are initially in the ground state, and $V_0 = 10\hbar^2/I$. Here, the states are labeled with a single running index $j$ and arranged in order of ascending energy, see panel (b) for the corresponding spectrum. Each band consists of 4 states. We observe that when the kick strength $\ell$ is an even number, the gears occupy only the non-degenerate states of highest and lowest energy in each band, which correspond to $k=0$ and $2$ and result in a vanishing time average of the relative momentum.

\begin{figure}
\centering
	\includegraphics[width=\columnwidth]{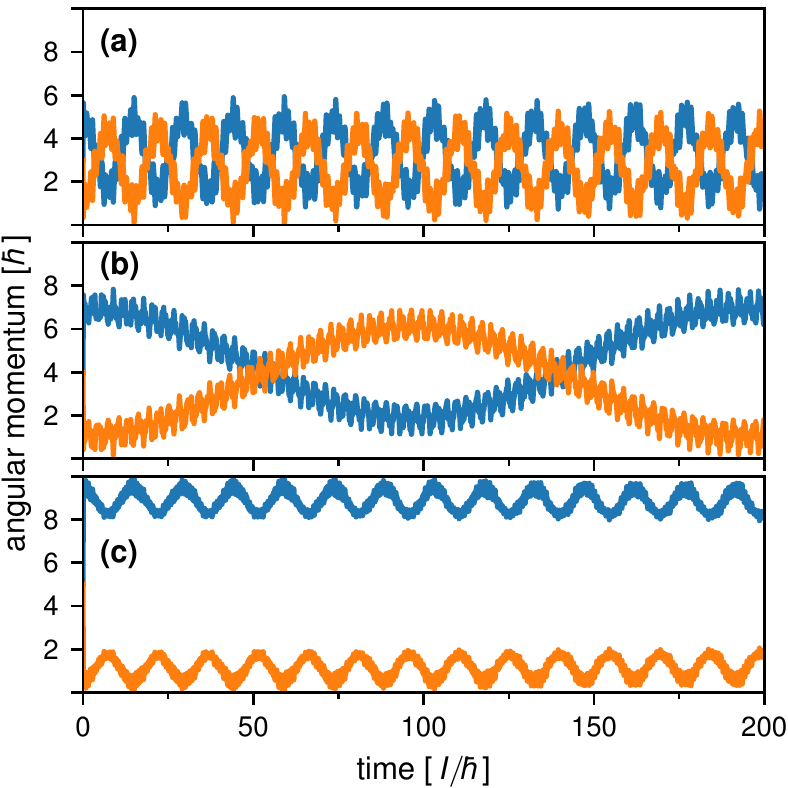}
\caption{Time evolution of the two gears' mean angular momenta $\la \oL_1\ra$ (blue) and $\la \oL_2 \ra$ (orange) after an initial kick of Gear 1 by $\ell=6$, 8, and 10 quanta, as plotted in (a), (b) and (c), respectively. We use the same parameters as in Fig.~\ref{fig:kick_energy_spec}.}\label{fig:single_kicked_evolution_2_2_10}
\end{figure}

However, it may indeed take a long time for the gears to reach the enhanced average transmission if the kick strength is large and excites the relative motion far beyond the interlocking threshold. Figure~\ref{fig:single_kicked_evolution_2_2_10}(a)-(c) shows the time evolution of the average angular momenta of Gear 1 (blue) and Gear 2 (orange) after a single kick of strength $\ell=6,8,10$, respectively.

In (a), the kick excites mostly the first four bands, and we find that the momenta of both gears oscillate with opposite phase around the benchmark long-time average $\overline{L_{1,2}} = 3\hbar$. The oscillation period, roughly $15 I/\hbar$, exceeds the inverse of the characteristic trapping frequency $\omega_0 = \ntot \sqrt{V_0/I_r}$ of the interlocking potential by an order of magnitude, here $2\pi/\omega_0 \approx 0.7 I/\hbar$.  Notice that the quantum revival time for the center-of-mass rotation is $\tau_c=8\pi I/\hbar$. 

The slightly stronger kick by $\ell=8$ in (b) leads to a significantly longer averaging period of about $200I/\hbar$, while in (c) the oscillation around the long-time average of $5\hbar$ can no longer be seen in the plotted time range. We can estimate the averaging period for the enhanced transmission by finding the two energy eigenvalues with the highest occupancy after the kick, see Fig.~\ref{fig:kick_energy_spec}(a), and taking the inverse of their difference. For the present case, this amounts to $15 I/\hbar$, $194 I/\hbar$, $4836 I/\hbar$, and $1.9 \times 10^5 I/\hbar$ for $\ell=6,8,10$, and $12$, respectively.

The plots also suggest that, if the coupling $V_0$ between the two gears could be switched off in a controlled manner, it is possible to transmit more angular momentum to Gear 2 than the benchmark average.

\subsection{Transmission in terms of ergotropy}\label{sec:sub_ergotropy}

So far we have described the transmission of motion between the quantum gears in terms of their mean angular momenta. Specifically, we quantified the transmission of an external kick on Gear 1 by the the fraction of angular momentum that ends up in Gear 2 \emph{on average} in the long-time limit. In the autonomous scenario where the coupling between the gears is constantly present, we found an enhanced transmission for certain kick strengths, but with the caveat that Gear 2 may require a long time to speed up and, eventually, it keeps oscillating around the predicted average velocity. 

\begin{figure}
\centering
\includegraphics[width=\columnwidth]{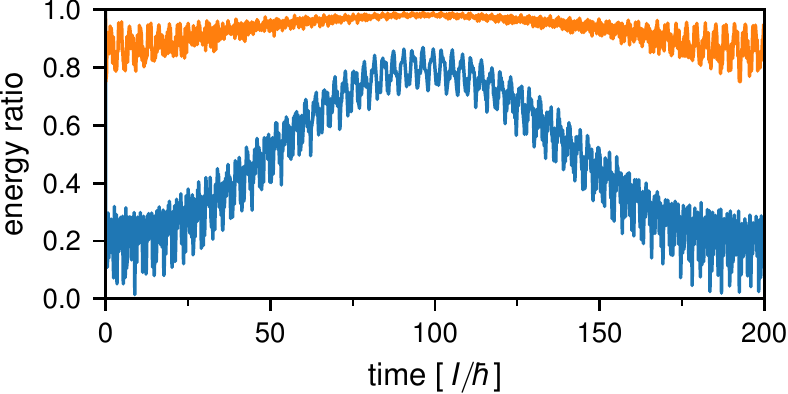}
\caption{Time evolution of the directional kinetic energy $\la \oL_2\ra^2/2I$ (blue) and the ergotropy $\mathcal{E}(\rho_2)$ (orange) contained in the motion of Gear 2, normalized to the kinetic energy $\la \oL_2^2 \ra/2I$. We consider an initial kick by $6$ quanta as in Fig.~\ref{fig:single_kicked_evolution_2_2_10}(b).}
\label{fig:ergotropy_2_3}
\end{figure}

Given that the time scales associated to the interlocking between the gears are much shorter than those transmission times, one might expect a faster transmission of energy than what is reflected in the net angular momentum of Gear 2. In fact, this discrepancy shows up in the amount of kinetic energy stored in Gear 2 \emph{on top of} the net directed motion it receives \cite{Seah2018,Seah2018a}. Figure~\ref{fig:ergotropy_2_3} shows the time evolution of the ratio between net kinetic energy $\la \oL_2 (t) \ra^2/2I$ and kinetic energy $\la \oL^2_2 (t) \ra/2I$ (blue line) for an exemplary scenario with enhanced transmission. Indeed, Gear 2 gains kinetic energy more quickly than it accelerates initially---but how much of that excess energy is useful, how much merely entropic?

This question is answered with the so-called ergotropy \cite{Allahverdyan2004,Goold2016}, an operational quantity that characterizes the useful energy stored in a quantum state by the amount of work that an external agent can at most extract in a cyclic unitary process. Here, we assume that the agent may only access Gear 2, i.e.~the ergotropy is defined in terms of its reduced quantum state $\rho_2$ and its local kinetic energy as
\begin{eqnarray}
\mathcal{E}(\rho_2) &=& \max_{\oU} \tr\left\{ \frac{\oL_2^2}{2I_2}\left(\rho_2-\oU \rho_2 \oU^\da \right) \right\} \nonumber \\
&=& \tr\left\{ \frac{\oL_2^2}{2I_2}\left(\rho_2-\rho_{\rm pa} \right) \right\}.\label{eqn:ergotropy_definition}
\end{eqnarray}
The unitary that maximizes this expression defines the (zero-ergotropy) passive state $\rho_{\rm pa}$ of Gear 2. It can be obtained from $\rho_2$ by taking its eigenspectrum of probabilities in descending order, $\{p_i | \, p_{i+1} \leq p_i, \, i \in \mathbb{N}_0 \}$, and redistributing it over the spectrum of momentum eigenstates $\{|m\ra \}$ in such a way that states of higher kinetic energy are less probable, 
\begin{equation}
\rho_{\rm pa} = p_0 |0\ra\la 0| + \sum_{m=1}^{\infty} \left( p_{2m-1} |m\ra \la m| + p_{2m}|-m \ra \la -m| \right).
\end{equation}
The kinetic energy content of this state is entirely of disordered, entropic nature.

In Figure~\ref{fig:ergotropy_2_3}, the ratio of ergotropy to kinetic energy is represented by the orange line, which stays close to unity and above the blue line. This indicates that Gear 2 acquires useful energy before it gains net momentum, and that almost all of the kinetic energy stored in Gear 2 is useful in the sense that it can be extracted as work. Note that in general, the ergotropy is an upper bound for the net kinetic energy, as one can extract the latter (up to an $\hbar$-fraction of residual momentum) by applying a unitary kick operator \eqref{eq:kickOperator}.

\subsection{Transmission of multiple kicks}
\label{sec:sub_multiple_kicks}

Finally, we consider the scenario where Gear 1 is driven by a sequence of multiple kick events rather than a single one. We assume that a total angular momentum of $\ell \hbar$ is supplied one quantum at a time in $\ell$ steps, with a waiting time $\Delta t$ in between. 

In a classical gear model, the waiting time $\Delta t$ will have a decisive influence on the overall transmission that can be achieved before the interlocking breaks. In fact, by synchronizing the kick sequence with the oscillatory motion in the relative two-gear coordinate, one can minimize the excitation of the oscillation amplitude and keep the gears interlocked.

In the quantum case of two gears initially prepared in the interlocked ground state, we find once again that an enhanced transmission could be achieved regardless of $\Delta t$ or  interlocking depth $V_0$. 
As in the single-kick case, the average transmission ratio of angular momentum assumes its benchmark value \eqref{eq:transmRatioBench} whenever the total number $\ell$ of kicks is a multiple of $ (n_1^2 + n_2^2)/2n_1$. The reason is that the time evolution by $\Delta t$ in between subsequent kicks preserves the Bloch wave number $k$ of the relative state, and so the previous reasoning in Sec.~\ref{sec:sub_single_kick} also applies to the cumulative change in relative momentum after $\ell$ steps.

\begin{figure}
\centering
\includegraphics[width=\columnwidth]{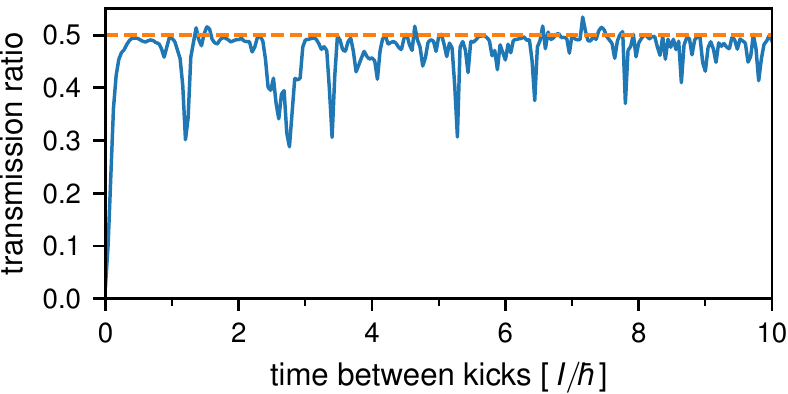}
\caption{Average transmission ratio after a sequence of 13 single-$\hbar$ kicks on Gear 1 with varying time between the kicks. We use identical gears with $n_{1,2}=2$ and $V_0 = 10 \hbar^2/I$. The orange dashed line marks the benchmark transmission for ideal classical gears.}
\label{fig:multiple_kicks}
\end{figure}

For all other $\ell$-values, the time-averaged transmission ratio \eqref{eq:transmRatio} varies strongly with the waiting time $\Delta t$, but the benchmark value \eqref{eq:transmRatioBench} would still be reached for specific $\Delta t$-values. This is illustrated in Fig.~\ref{fig:multiple_kicks} for identical gears with two teeth and an odd number of steps, $\ell=13$. There we observe that the transmission ratio stays close to the benchmark value $0.5$ if the waiting time between successive kicks is long, while it can fall significantly below $0.5$ at shorter times.

\section{Conclusion}

We have studied a model for quantum gears based on coherently interacting planar rotors. Focusing on the basic scenario of two such gears and an external source that injects angular momentum into one of them, we could identify several observable quantum features in their behavior resulting from the angular symmetries and the quantization of angular momentum.

The combined gear motion separates into a free rotor degree of freedom that describes the synchronous rotation of the interlocked gears and the bound relative motion of the teeth in the periodic interlocking potential. The former exhibits quantum state revivals at a fixed period of time that scales with the overall mass of the gears, while the latter results in a band structure and Bloch wave phenomena. Specifically, when the gears are initially prepared in their perfectly interlocked ground state, we find that the average transmission of external angular momentum from one gear to the other can either fall below the classical prediction for ideal hard gears due to tunneling, or it can reach that ideal value even the interlocking breaks due to constructive interference. Finally, by assessing the transmission of motion in terms of ergotropy rather than net angular momentum, we find that useful mechanical energy transmits faster than directed motion. 

Future studies could address
realistic nanomechanical implementations, 
damping and diffusion caused by thermal reservoirs \cite{Stickler2018}, time-dependent driving by external control fields, and coupling to other optomechanical degrees of freedom.

\begin{acknowledgments}
We thank Alexandre Roulet, Benjamin Stickler, and Christoph Bruder for helpful discussions.
This research is supported by the Singapore Ministry of Education through the Academic Research Fund Tier 3 (Grant No. MOE2012-T3-1-009); and by the same MoE and the National Research Foundation, Prime Minister's Office, Singapore, under the Research Centres of Excellence programme.
\end{acknowledgments}


%

\end{document}